\documentclass[12pt]{iopart}

\usepackage{iopams}
\usepackage[english]{babel}
\usepackage{graphicx}
\usepackage{cite}
\usepackage{bbm}
\usepackage{mathenv}
\usepackage{color}
\usepackage{float}
\usepackage{braket}
\begin{document}

\title{Purification Hamiltonians}

\author{Jorge A. Anaya-Contreras,$^1$ Arturo Z\'u\~niga-Segundo$^1$ and H\'ector M. Moya-Cessa$^2$}

\address{$^1$Instituto Polit\'ecnico Nacional, ESFM, Departamento de F\'isica. Edificio 9, Unidad Profesional ``Adolfo L\'opez Mateos,'' CP 07738 CDMX, Mexico\\ $^2$Instituto Nacional de Astrof\'{\i}sica, \'Optica y Electr\'onica INAOE\\ Calle Luis Enrique Erro 1, Santa Mar\'{\i}a Tonantzintla, Puebla, 72840 Mexico}

\begin{abstract}
We study the Jaynes-Cummings interaction when initial mixed states, for the atom or the field, are considered. The evolved mixed field density matrix is purified to a wavefunction that describes the interaction between a quantised field and an artificial four-level atom. This allows us to  use the Araki-Lieb inequality  to calculate the field entropy from the atomic entropy. We then generate {\it artificial} Hamiltonians that reproduce the field entropy dynamics. We finally show that more realistic Hamiltonians may be considered by using  two entangled two-level atoms.
\end{abstract}



\maketitle

\maketitle

\section{Introduction}
The atomic inversion for a two-level atom interacting with a quantised \cite{Jaynes} field undergoes collapses and revivals of Rabi oscillations for several initial field states  \cite{Shore,Gerry}. It is well-known that such revivals are an indicator of the nature of the quantised field  inside the cavity because the atomic inversion depends on the photon number distribution. As an example, if a squeezed field is considered, the atomic inversion suffers so-called ringing revivals that give us information that  specific non-classical field was used as an initial state \cite{Satya,Vidiella}. 

The von Neumann entropy \cite{Neumann},  together with the atomic inversion may give information about the generation of nonclassical states, as the first tell us about the degree of purity of the state while the second, as already mentioned, points out which photon distribution is used. The entropy may  help to decide if a coherent state \cite{Glauber} or an statistical mixture of them were used, as they both  produce the same atomic inversion \cite{Banacloche,Phoenix2}.

Although the present contribution is devoted to the atom-field interaction, because of the similarity with other systems such as ion-laser interactions \cite{vogeli,vogelii,Leibfried,Jonathan} or the propagation of light through waveguide arrays \cite{Leija,Lara}, the results obtained here are also valid  for such interactions.

One of the main tasks in the present manuscript is to calculate the entropy of the field, which we will do with the aid of the Araki-Lieb inequality \cite{Araki}
\begin{equation}
\mid S_A-S_F\mid\le S_{AF}\le S_A+S_F
\end{equation}
where $S_{AF}$ is the von Neumann entropy  of the composed system, atom and field,   $S_A$ and $S_F$ are the reduced entropies for the atom and the field, respectively. This inequality will help us to calculate the entropy of the subsystems if both of them were initially in pure states. 

Because we will consider mixtures as initial states \cite{Zuniga}, in principle it will not be possible to use the Araki-Lieb inequality to calculate the entropies, especially the field entropy. However via purification of the mixed density matrix of the qunatised field \cite{Araki,Pathak,Anaya}, we will be able to use the Araki-Lieb inequality in order calculate the field von Neumann entropy even in the case of initial statistical mixtures, either for the atom or the field.

The purification will allow us to define different interaction Hamiltonians (some artificial, some more realistic) that reproduce the same field entropies. 

In next Section we study the atom-field interaction and define the initial mixed states for the atom and the field that lead us to different interaction Hamiltonians. Section III treats the case of two two-level atoms in an experimentally feasible configuration  that models the Jaynes-Cummings interaction with mixed states and reproduces the same field entropies. Section IV is left for conclusions.

\section{Atom-field interaction}
First we look at the atom field interaction. We will pay particular attention to initial conditions where atom or field are initially in mixed states such that the Araki-Lieb inequality can not be applied directly. We use the well-known Hamiltonian for a two-level atom interacting with a quantised field under the rotating wave approximation (we set $\hbar=1$)
\begin{equation}
\hat{H} = \omega\hat{a}^\dagger \hat{a} +  \frac{\omega_{A}}{2}\hat{\sigma}_{z}
+ \lambda(\hat{a}\sigma^{+}+\hat{a}^\dagger \sigma^{-}),
\label{1}
\end{equation}
i.e., the Jaynes-Cummmings model Hamiltonian. In the equation above, the operators  $\hat{a}$  and  $\sigma^{-}$ are the annihilation operators of the quantised field and
the two-level atom flip operator (one of the Pauli spin matrices) for the $|e\rangle \rightarrow |g\rangle$  transition of frequency $\omega_{A}$,
respectively. The field frequency is $\omega$  and  $\lambda$ is the atom-field  coupling constant.
The interaction Hamiltonian, after getting rid off the free terms, has the form
\begin{equation}
\hat{H}_I =  \lambda(\hat{a}\sigma^{+}+\hat{a}^\dagger \sigma^{-}).
\label{1}
\end{equation}
The evolution operator, $\hat{U}_I(t)=e^{-i\hat{H}_It}$, reads (in the $2\times 2$ matrix representation, in fact we will pass from this representation to the Pauli spin operators throughout the manuscript)
\begin{equation}
\label{AV1}
\hat{U}_I(t)
=
\begin{pmatrix}
C_{\hat{n}+1} & -i S_{\hat{n}+1}\hat{V} \vspace{0.2cm}\\
-i \hat{V}^{\dagger}S_{\hat{n}+1}& C_{\hat{n}} 
\end{pmatrix}\,,
\end{equation}
where $C_{\hat{n}+1}=\cos(\lambda t \sqrt{\hat{n}+1})$, $S_{\hat{n}+1}=\sin(\lambda t \sqrt{\hat{n}+1})$ with $\hat{n}=\hat{a}^{\dagger}\hat{a}$. The operator $\hat{V}=\frac{1}{\sqrt{\hat{n}+1}}\hat{a}$ is the London phase operator \cite{London}.
\subsection{Initial state: atom in a pure state and field in an statistical mixture of coherent states}
We consider first the initial mixed state given by 
\begin{equation}
\hat{\rho}(0)=\left(\mathrm{C}\ket{\alpha}\bra{\alpha}+(1-\mathrm{C})\ket{\beta}\bra{\beta}\right)\ket{e}\bra{e},
\end{equation} 
where $C$  ensures normalization and $|\alpha\rangle$ ($|\beta\rangle$) is a  coherent states of amplitude $\alpha$ ($\beta$) \cite{Glauber}
\begin{equation}
\ket{\alpha}=e^{-\frac{|\alpha|^2}{2}}\sum_{n=0}^{\infty}\frac{\alpha^n}{\sqrt{n!}}\ket{n},
\end{equation} 
with $\ket{n}$ a number state.

By applying the evolution operator to this states, we obtain the evolved density matrix, $\hat{\rho}(t)=\hat{U}_I(t)\hat{\rho}(0)\hat{U}_I^{\dagger}(t)$, 
\begin{equation}
\label{AV3}
\hat{\rho}(t)=
\begin{pmatrix}
\ket{\psi_{1}(t)}\bra{\psi_{1}(t)}+\ket{\psi_{3}(t)}\bra{\psi_{3}(t)} & \ket{\psi_{1}(t)}\bra{\psi_{2}(t)}+\ket{\psi_{3}(t)}\bra{\psi_{4}(t)} \vspace{0.2cm}\\ 
\ket{\psi_{2}(t)}\bra{\psi_{1}(t)}+\ket{\psi_{4}(t)}\bra{\psi_{3}(t)} & \ket{\psi_{2}(t)}\bra{\psi_{2}(t)}+\ket{\psi_{4}(t)}\bra{\psi_{4}(t)}
\end{pmatrix}\,,
\end{equation}
with the unnormalized wavefunctions  given by
\begin{eqnarray}
\label{AV6}
\ket{\psi_{1}(t)}&=&\sqrt{\mathrm{C}}\cos(\lambda t \sqrt{\hat{a}\hat{a}^{\dagger}})\ket{\alpha}\,, \nonumber\\
\ket{\psi_{2}(t)}&=&-\,i\sqrt{\mathrm{C}}\,\hat{V}^{\dagger}\sin(\lambda t \sqrt{\hat{a}\hat{a}^{\dagger}})\ket{\alpha}\,, \nonumber\\
\ket{\psi_{3}(t)}&=&\sqrt{1-\mathrm{C}}\cos(\lambda t \sqrt{\hat{a}\hat{a}^{\dagger}})\ket{\beta}\,, \nonumber\\
\ket{\psi_{4}(t)}&=&-\,i\sqrt{1-\mathrm{C}}\,\hat{V}^{\dagger}\sin(\lambda t \sqrt{\hat{a}\hat{a}^{\dagger}})\ket{\beta}\,.
\end{eqnarray}
The reduced density matrices for atom and field are the written as
\begin{equation}
\label{AV4}
\hat{\rho}_{A}=
\begin{pmatrix}
\braket{\psi_{1}|\psi_{1}}+\braket{\psi_{3}|\psi_{3}} & \braket{\psi_{1}|\psi_{2}}^{*}+\braket{\psi_{3}|\psi_{4}}^{*}  \vspace{0.2cm}\\
\braket{\psi_{1}|\psi_{2}}+\braket{\psi_{3}|\psi_{4}} & \braket{\psi_{2}|\psi_{2}}+\braket{\psi_{4}|\psi_{4}}  
\end{pmatrix}\,,
\end{equation}
and
\begin{equation}
\label{AV5}
\hat{\rho}_{F}=\ket{\psi_{1}}\bra{\psi_{1}}+\ket{\psi_{2}}\bra{\psi_{2}}+\ket{\psi_{3}}\bra{\psi_{3}}+\ket{\psi_{4}}\bra{\psi_{4}}\,.
\end{equation}
respectively. Purification  \cite{Anaya} of the above state gives the wavefunction 
\begin{equation}
\label{AV7}
\ket{\psi }=\ket{\psi_{1}}\ket{A_{1}}+\ket{\psi_{2}}\ket{A_{2}}+\ket{\psi_{3}}\ket{A_{3}}+\ket{\psi_{4}}\ket{A_{4}} ,
\end{equation}
so that we have passed from a two-dimensional Hilbert space for the atom, given by the real states $|e\rangle$ and $|g\rangle$ to a four dimensional Hilbert space for an artificial atom, given by the states 
\begin{equation}
\label{AV8}
\ket{A_{1}}=\begin{pmatrix}1\\ 0\\0\\0 \end{pmatrix} \,,\,\,\,\, \ket{A_{2}}=\begin{pmatrix}0\\ 1\\0\\0 \end{pmatrix} \,, \,\,\,\, \ket{A_{3}}=\begin{pmatrix}0\\ 0\\1\\0 \end{pmatrix} \,,\,\,\,\, \ket{A_{4}}=\begin{pmatrix}0\\0\\0\\1 \end{pmatrix}\,.
\end{equation}
\subsection*{Evolution operator for the artificial atom}
There are a number of (artificial) evolution operators that we could imagine that would render equation (\ref{AV7}), among them, the one given by the $4\times 4$ matrix in the following equation
\begin{equation}
\label{AV9}
\ket{\psi }=
\begin{pmatrix}
C_{\hat{n}+1} & -\,i S_{\hat{n}+1}\hat{V} & 0 &0 \vspace{0.2cm} \\
-\,i\,\hat{V}^{\dagger}S_{\hat{n}+1} & C_{\hat{n}} & 0 & 0 \vspace{0.2cm} \\
0 & 0 & C_{\hat{n}+1} & -\,i S_{\hat{n}+1}\hat{V} \vspace{0.2cm} \\
0 & 0 & -\,i\,\hat{V}^{\dagger}S_{\hat{n}+1}& C_{\hat{n}}
\end{pmatrix}\begin{pmatrix}\sqrt{\mathrm{C}}\ket{\alpha}\\ 0 \\ \sqrt{1-\mathrm{C}}\ket{\beta} \\ 0 \end{pmatrix}\,,
\end{equation}
with the artificial initial pure state given by
\begin{equation}
\label{AV10}
\ket{\psi (0)}=\begin{pmatrix}\sqrt{\mathrm{C}}\ket{\alpha}\\ 0 \\ \sqrt{1-\mathrm{C}}\ket{\beta} \\ 0 \end{pmatrix} = \sqrt{\mathrm{C}}\ket{\alpha}\ket{A_{1}}+ \sqrt{1-\mathrm{C}}\ket{\beta}\ket{A_{3}}\,.
\end{equation}
Because the evolution operator is the exponential of the Hamiltonian, namely
\begin{equation}
\label{AV11}
\hat{U} =\exp(-it\hat{H} )\end{equation}
we can determine the artificial Hamiltonian that produces the purified state (\ref{AV7}), that would represent the interaction between the artificial four-level atom and the quantised field
\begin{eqnarray}
\label{AV12}
\hat{H} = \lambda \hat{a}(\ket{A_{1}}\bra{A_{2}}+\ket{A_{3}}\bra{A_{4}})+\lambda \hat{a}^{\dagger}(\ket{A_{2}}\bra{A_{1}}+\ket{A_{4}}\bra{A_{3}})\,.
\end{eqnarray}
\subsubsection{Atom and field entropies}
The  von Neumann entropy \cite{Neumann} for the a density matrix,  $\hat{\rho}$, is defined as  
\begin{equation}
\label{AVE1}
S=-\Tr\left\lbrace\hat{\rho}\ln\hat{\rho}\right\rbrace\,.
\end{equation}
\section*{Entropies}For the initial mixed state $ 
\hat{\rho}(0)=\left(\mathrm{C}\ket{\alpha}\bra{\alpha}+(1-\mathrm{C})\ket{\beta}\bra{\beta}\right)\ket{e}\bra{e}$ it is easy to find the  it is easy to find the atomic entropy 
\begin{equation}
\label{AVE2}
S_{A}=-\Lambda_{1}\ln\Lambda_{1}-\Lambda_{2}\ln\Lambda_{2}
\end{equation}
and the field entropy, because the initial state in Equation (13) is in a pure state, we can obtain it from the four-level atom's entropy
\begin{equation}
\label{AVE3}
S_{F}=-\lambda_{1}\ln\lambda_{1}-\lambda_{2}\ln\lambda_{2}-\lambda_{3}\ln\lambda_{3}-\lambda_{4}\ln\lambda_{4} \,.
\end{equation}
In the above equations, $\Lambda_{j},\,\,  j=1,2$, are the eigenvalues of the matrix (\ref{AV4}) and $\lambda_{j},\,\, j=1, \dots, 4$, eigenvalues of the matrix $\hat{\rho}_{AA}$ (artificial atom)
\begin{equation}
\label{AVE5}
\hat{\rho}_{AA}=
\begin{pmatrix}
\mathrm{P}_{11} & \mathrm{P}^{*}_{12} & \mathrm{P}^{*}_{13} & \mathrm{P}^{*}_{14} \vspace{0.2cm}\\
\mathrm{P}_{12} & \mathrm{P}_{22} & \mathrm{P}^{*}_{23} & \mathrm{P}^{*}_{24} \vspace{0.2cm}\\
\mathrm{P}_{13} & \mathrm{P}_{23} & \mathrm{P}_{33} & \mathrm{P}^{*}_{34} \vspace{0.2cm}\\
\mathrm{P}_{14} & \mathrm{P}_{24} & \mathrm{P}_{34} & \mathrm{P}_{44} 
\end{pmatrix}\,,
\end{equation} 
with $\mathrm{P}_{ij}=\braket{\psi_{i}|\psi_{j}}$ for all $i,j \in\{1,2,3,4\}$.

\subsection{Initial state: field in pure state and atom  in statistical mixture }
In this subsection we will basically produce the same equations as in the former, except for the fact that the different initial condition will produce some subtle differences. For the initial condition 
\begin{equation}
\hat{\rho}(0)=\left(\mathrm{C}\ket{e}\bra{e}+(1-\mathrm{C})\ket{g}\bra{g}\right)\ket{\alpha}\bra{\alpha}\,,
\end{equation}
 we obtain the evolved density matrix
\begin{equation}
\label{AV14}
\hat{\rho}=
\begin{pmatrix}
\ket{\Psi_{1}(t)}\bra{\Psi_{1}(t)}+\ket{\Psi_{3}(t)}\bra{\Psi_{3}(t)} & \ket{\Psi_{1}(t)}\bra{\Psi_{2}(t)}+\ket{\Psi_{3}(t)}\bra{\Psi_{4}(t)} \vspace{0.2cm}\\ 
\ket{\Psi_{2}(t)}\bra{\Psi_{1}(t)}+\ket{\Psi_{4}(t)}\bra{\Psi_{3}(t)} & \ket{\Psi_{2}(t)}\bra{\Psi_{2}(t)}+\ket{\Psi_{4}(t)}\bra{\Psi_{4}(t)}
\end{pmatrix}\,,
\end{equation}
where now the unnormalized wavefunctions are given by
\begin{eqnarray}
\label{AV17}
\ket{\Psi_{1}(t)}&=&\sqrt{\mathrm{C}}\cos(\lambda t \sqrt{\hat{a}\hat{a}^{\dagger}})\ket{\alpha}\,, \nonumber\\
\ket{\Psi_{2}(t)}&=&-\,i\sqrt{\mathrm{C}}\,\hat{V}^{\dagger}\sin(\lambda t \sqrt{\hat{a}\hat{a}^{\dagger}})\ket{\alpha}\,, \nonumber\\
\ket{\Psi_{3}(t)}&=&-\,i\sqrt{1-\mathrm{C}}\,\sin(\lambda t \sqrt{\hat{a}\hat{a}^{\dagger}})\hat{V}\ket{\alpha}\,, \nonumber\\
\ket{\Psi_{4}(t)}&=&\sqrt{1-\mathrm{C}}\,\cos(\lambda t \sqrt{\hat{a}^{\dagger}\hat{a}})\ket{\alpha}\,,
\end{eqnarray} 
and the differences with the former case may be noted in the forms of $\ket{\Psi_{3}}$ and $\ket{\Psi_{4}}$. The reduced density operators take the same form as before
\begin{equation}
\label{AV15}
\hat{\rho}_{A}=
\begin{pmatrix}
\braket{\Psi_{1}|\Psi_{1}}+\braket{\Psi_{3}|\Psi_{3}} & \braket{\Psi_{1}|\Psi_{2}}^{*}+\braket{\Psi_{3}|\Psi_{4}}^{*}  \vspace{0.2cm}\\
\braket{\Psi_{1}|\Psi_{2}}+\braket{\Psi_{3}|\Psi_{4}} & \braket{\Psi_{2}|\Psi_{2}}+\braket{\Psi_{4}|\Psi_{4}}  
\end{pmatrix}\,,
\end{equation}
\begin{equation}
\label{AV16}
\hat{\rho}_{F}=\ket{\Psi_{1}}\bra{\Psi_{1}}+\ket{\Psi_{2}}\bra{\Psi_{2}}+\ket{\Psi_{3}}\bra{\Psi_{3}}+\ket{\Psi_{4}}\bra{\Psi_{4}}\,,
\end{equation}
for the atom and field, respectively.

Purification  of the state (\ref{AV16}) gives the wavefunction
\begin{equation}
\label{AV18}
\ket{\Psi }=\ket{\Psi_{1}}\ket{A_{1}}+\ket{\Psi_{2}}\ket{A_{2}}+\ket{\Psi_{3}}\ket{A_{3}}+\ket{\Psi_{4}}\ket{A_{4}} 
\end{equation}
that may be obtained now from the initial state
\begin{equation}
\label{AV21}
\ket{\psi (0)}= \begin{pmatrix}\sqrt{\mathrm{C}}\ket{\alpha}\\ 0 \\ 0 \\ \sqrt{1-\mathrm{C}}\ket{\alpha} \end{pmatrix} = (\sqrt{\mathrm{C}}\ket{A_{1}}+\sqrt{1-\mathrm{C}}\ket{A_{4}})\ket{\alpha}\,.
\end{equation}
with the evolution operator
\begin{equation}
\label{AV22}
\hat{U} =
\begin{pmatrix}
C_{\hat{n}+1}& -\,i S_{\hat{n}+1}\hat{V} & 0 &0 \vspace{0.2cm} \\
-\,i\,\hat{V}^{\dagger}S_{\hat{n}+1} & C_{\hat{n}} & 0 & 0 \vspace{0.2cm} \\
0 & 0 & C_{\hat{n}+1} & -\,i S_{\hat{n}+1}\hat{V} \vspace{0.2cm} \\
0 & 0 & -\,i\,\hat{V}^{\dagger}S_{\hat{n}+1} & C_{\hat{n}}
\end{pmatrix}\,,
\end{equation}
and, therefore, the artificial Hamiltonian given as before
\begin{equation}
\label{AV23}
\hat{H} =\lambda \hat{a}(\ket{A_{1}}\bra{A_{2}}+\ket{A_{3}}\bra{A_{4}})+\lambda \hat{a}^{\dagger}(\ket{A_{2}}\bra{A_{1}}+\ket{A_{4}}\bra{A_{3}})\,.
\end{equation}

\begin{figure}[h!]
\centering
\includegraphics[width=8cm]{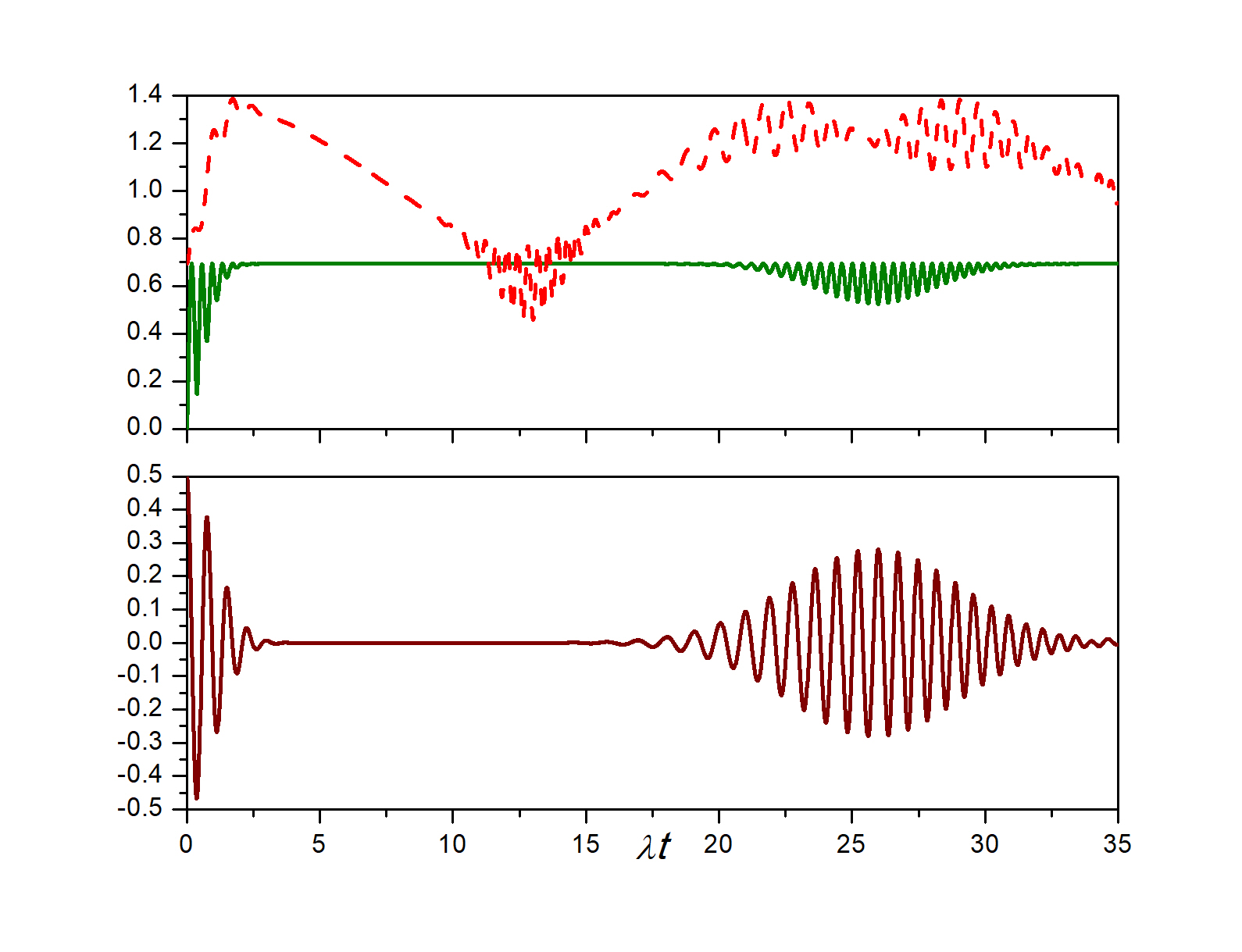} 
\caption{The figure (above) shows the atomic (solid line) and field (dashed line) entropies for the initial state $\hat{\rho}(0)=\left(\mathrm{C}\ket{\alpha}\bra{\alpha}+(1-\mathrm{C})\ket{\beta}\bra{\beta}\right)\ket{e}\bra{e}$, with $\mathrm{C}=0$.5 and $\alpha=-\beta=4.0.$ Below we plot the atomic inversion $W(t)=P_e(t)-P_g(t)$ where $P_e(t)$ ($P_g(t)$) is the probability to find the two-level atom in its excited (ground) state.}
\label{(FVAMS2LJCM2FS)PSFSAC05A4BM4}
\end{figure}
\begin{figure}[h!]
\centering
\includegraphics[width=8cm]{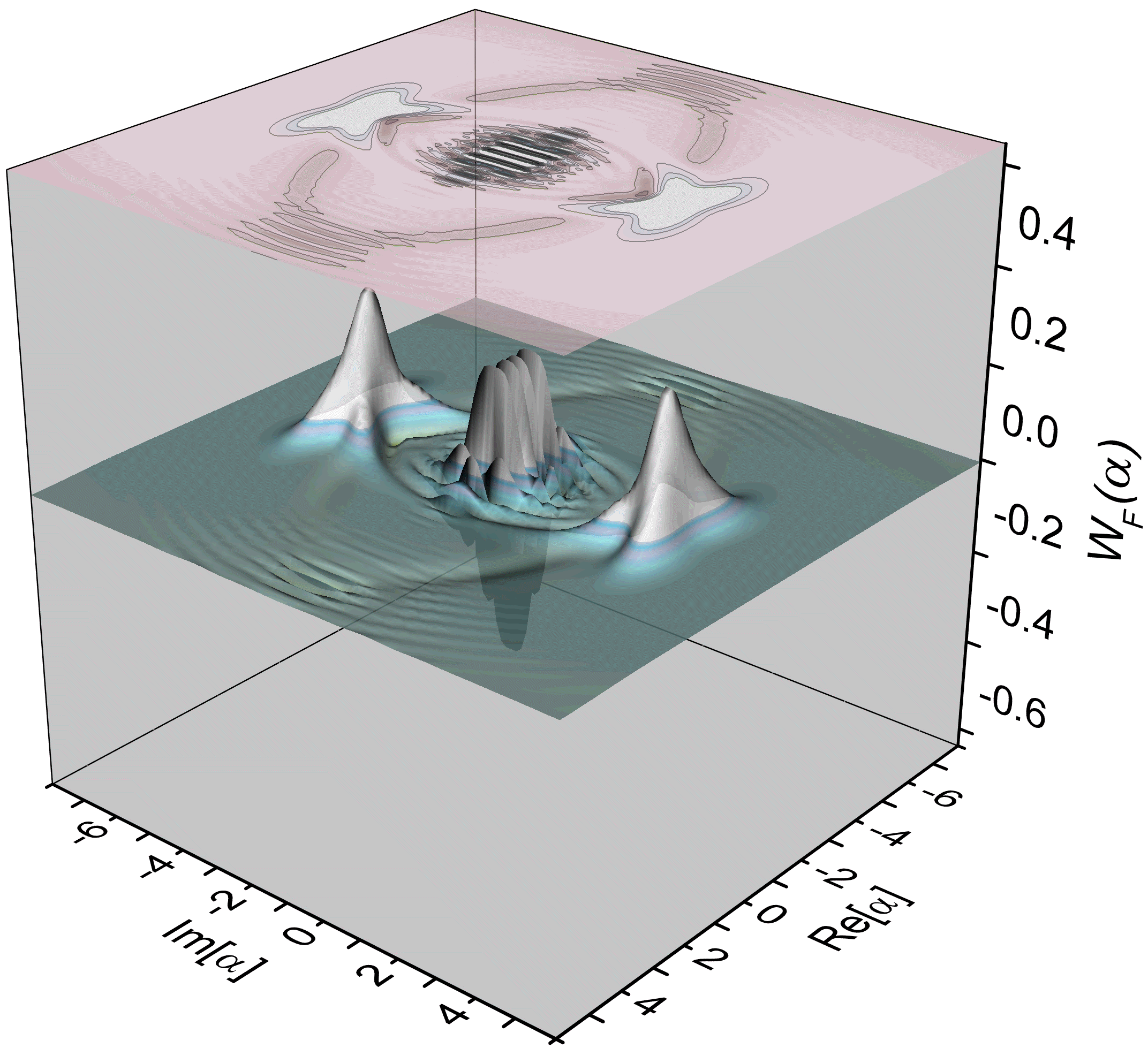} 
\caption{We plot the Wigner fucntion for the state given in Fig. 1 at a time $t\approx 12.54/\lambda$.}
\end{figure}
\subsection*{Entropies} For the initial mixed state $\hat{\rho}(0)=\left(\mathrm{C}\ket{e}\bra{e}+(1-\mathrm{C})\ket{g}\bra{g}\right)\ket{\alpha}\bra{\alpha}\,$, the von Newmann entropies associated to the field and the atom are
\begin{equation}
\label{AVE8}
S_{A}=-\Lambda_{1}\ln\Lambda_{1}-\Lambda_{2}\ln\Lambda_{2}
\end{equation}
and
\begin{equation}
\label{AVE9}
S_{F}=-\lambda_{1}\ln\lambda_{1}-\lambda_{2}\ln\lambda_{2}-\lambda_{3}\ln\lambda_{3}-\lambda_{4}\ln\lambda_{4} \,,
\end{equation}
respectively. The  $\Lambda_{j}\,\, j={1,2}$ correspond to the eigenvalues of the atomic density matrix (\ref{AV15}) and $\lambda_{j}\,\,  j=1\dots 4$ are the eigenvalues of $\hat{\rho}_{AA}$ (the artificial atom)
\begin{equation}
\label{AVE10}
\hat{\rho}_{AA}=
\begin{pmatrix}
\mathrm{P}_{11} & \mathrm{P}^{*}_{12} & \mathrm{P}^{*}_{13} & \mathrm{P}^{*}_{14} \vspace{0.2cm}\\
\mathrm{P}_{12} & \mathrm{P}_{22} & \mathrm{P}^{*}_{23} & \mathrm{P}^{*}_{24} \vspace{0.2cm}\\
\mathrm{P}_{13} & \mathrm{P}_{23} & \mathrm{P}_{33} & \mathrm{P}^{*}_{34} \vspace{0.2cm}\\
\mathrm{P}_{14} & \mathrm{P}_{24} & \mathrm{P}_{34} & \mathrm{P}_{44} 
\end{pmatrix}\,,
\end{equation} 
with $\mathrm{P}_{ij}=\braket{\Psi_{i}|\Psi_{j}}$ for each $i,j \in\{1,2,3,4\}$.

\begin{figure}[h!]
\centering
\includegraphics[width=8cm]{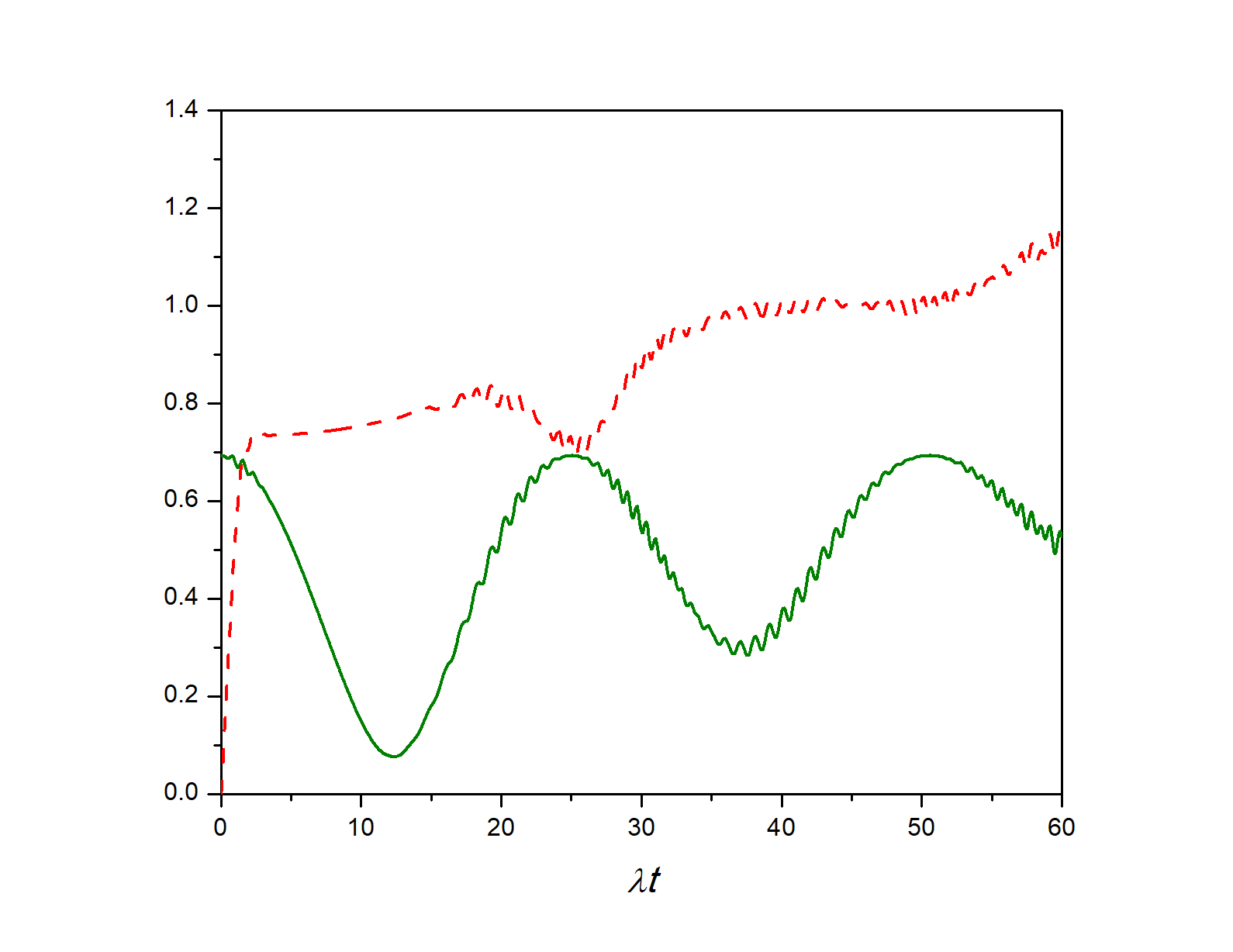} 
\caption{We plot the atomic (solid line) and field (dashed line) entropies for the initial state $\hat{\rho}(0)=\left(\mathrm{C}\ket{e}\bra{e}+(1-\mathrm{C})\ket{g}\bra{g}\right)\ket{\alpha}\bra{\alpha}$, with $\mathrm{C}=0$.5 and $\alpha=4$.0.}
\label{(FVAMS2LJCM2AS)PSFSAC05A4}
\end{figure}

From Fig. \ref{(FVAMS2LJCM2FS)PSFSAC05A4BM4} we find that the field entropy takes the initial value $\ln2$ because the field was prepared in an statistical mixture of coherent states. On the other hand, in figure (\ref{(FVAMS2LJCM2AS)PSFSAC05A4}) the field entropy is zero for $t=0$ as the field was prepared in a pure state. It may also be seen that the field becomes purer at some times after evolution.

In  Fig. 1 (below), we plot the atomic inversion as a function of time, 
\begin{equation}
W(t)=\sum_{n=0}^{\infty}P_n\cos(\lambda t\sqrt{n+1}).
\end{equation}
We may note  the usual behaviour of it: a collapse region that is produced by the initial mixture of two  coherent states (as it has the same photon distribution of a coherent state). The initial coherent states located at $x=\alpha$ and $x=-\alpha$ separate in two spots each, travelling in opposite directions (clockwise and anticlockwise). As they collide at $x=0$ in phase space (this may be seen in Fig. 2, where we plot the Wigner function \cite{Wigner}) there is no effect in the atomic inversion. However, this collision may be noted in the oscillations presented by the field entropy (Fig. 1 - above part). 

We want to stress that, to the best of our knowledge, effects of the collision of coherent states that form an statistical mixture had not been shown before.

Fig. 3 shows the atomic and field entropies for the atom initially in a mixed state and the field in a coherent state, Equation (21). It may be seen that the atom starts initially maximally mixed to become purer after evolution while the field entropy grows towards the value $\ln 4$, predicted by the state in Equation (26). 

In order to conclude this Section we may say that the Jaynes-Cummings model with different initial conditions, for instance the  {\it non-pure} states $\hat{\rho}(0)=\left(\mathrm{C}\ket{\alpha}\bra{\alpha}+(1-\mathrm{C})\ket{\beta}\bra{\beta}\right)\ket{e}\bra{e}$, or $\hat{\rho}(0)=\left(\mathrm{C}\ket{e}\bra{e}+(1-\mathrm{C})\ket{g}\bra{g}\right)\ket{\alpha}\bra{\alpha}$   may be mimicked by the interaction between an artificial four-level atom and a quantised field with initial conditions given by {\it pure} entangled states $\ket{\psi (0)} = \sqrt{\mathrm{C}}\ket{\alpha}\ket{A_{1}}+ \sqrt{1-\mathrm{C}}\ket{\beta}\ket{A_{3}}$  and {\it pure} non-entanglend state $\ket{\psi (0)}= (\sqrt{\mathrm{C}}\ket{A_{1}}+\sqrt{1-\mathrm{C}}\ket{A_{4}})\ket{\alpha}$, respectively.

\section{Two-atom Hamiltonian leading to purification}
Consider now the Hamiltonian given in Equation (1) but with an extra atom outside the cavity
\begin{equation}
\hat{H} = \omega\hat{a}^{\dagger} \hat{a} +  \frac{\omega_{A}}{2}\hat{\sigma}_{z,1}
+ \lambda[\hat{a}\sigma_1^{+}+\hat{a}^{\dagger} \sigma_1^{-}]+ \frac{\omega_{a}}{2}\hat{\sigma}_{z,2},
\label{twoatoms}
\end{equation}
where $\omega_a$ is the second atom's atomic transition frequency. The Pauli spin operators have been indexed 1 and 2 for the different atoms.

By going to the interaction picture, {\it  i.e.}, getting rid off the free Hamiltonians, we can obtain the evolved state from the initial state \cite{expl1}
\begin{equation}
|\psi(0)\rangle =\frac{1}{\sqrt{2}}|\alpha\rangle (|e_1\rangle|e_2\rangle+|g_1\rangle|g_2\rangle)
\end{equation}
as
\begin{eqnarray}
|\psi(t)\rangle &=&\frac{1}{\sqrt{2}}\left(\cos(\lambda  t \sqrt{\hat{n}+1})|\alpha\rangle |e_1\rangle|e_2\rangle-i\hat{V}^{\dagger}\sin(\lambda  t \sqrt{\hat{n}+1})|\alpha\rangle |g_1\rangle|e_2\rangle\right)\\ \nonumber &-&\frac{1}{\sqrt{2}} \left(i\sin(\lambda  t \sqrt{\hat{n}+1})\hat{V}|\alpha\rangle |e_1\rangle|g_2\rangle+
\cos(\lambda  t \sqrt{\hat{n}})|\alpha\rangle |g_1\rangle|g_2\rangle\right).
\end{eqnarray}
Note that the initial state (35) has an atom-atom entangled state while the field is initially in a coherent state. 

By identifying $|e_1\rangle|e_2\rangle\rightarrow |A_1\rangle$, $|e_1\rangle|e_2\rangle\rightarrow |A_2\rangle$, $|g_1\rangle|e_2\rangle\rightarrow |A_3\rangle$
and $|g_1\rangle|g_2\rangle\rightarrow |A_4\rangle$, we recover the {\it purified} state (23)  for $C=0.5$.

If instead we use an initial state of the form \cite{expl2}
\begin{equation}
|\psi(0)\rangle =\frac{1}{\sqrt{2}}|e_1\rangle (|\alpha\rangle|e_2\rangle+|-\alpha\rangle|g_2\rangle)
\end{equation}
i.e., now we consider atom one in its excited state and atom two entangled with the field, we obtain the evolved wavefunction as
\begin{eqnarray}
|\psi(t)\rangle &=&\frac{1}{\sqrt{2}}\left(\cos(\lambda  t \sqrt{\hat{n}+1})|\alpha\rangle |e_1\rangle|e_2\rangle-i\hat{V}^{\dagger}\sin(\lambda  t \sqrt{\hat{n}+1})|\alpha\rangle |g_1\rangle|e_2\rangle\right)\\ \nonumber &+& \frac{1}{\sqrt{2}}\left(\cos(\lambda  t \sqrt{\hat{n}+1})|-\alpha\rangle |e_1\rangle|g_2\rangle-i\hat{V}^{\dagger}\sin(\lambda  t \sqrt{\hat{n}+1})|-\alpha\rangle |g_1\rangle|g_2\rangle \right)\,
\end{eqnarray}
that, together with the atomic states identified above,  give the purified state (11).
\section{Conclusions}
We have shown that the Jaynes-Cummings interaction when initial mixed states are chosen may be modelled by some artificial Hamiltonians.  In particular, we have seen that the purification process that takes us from a mixed field density matrix to a pure wave function that involves a four-level system, could be generated either by the interaction between the field and  a four-level atom or the field with a single  two-level  atom with special entangled initial conditions with a second atom outside the cavity.  Finally we should mention that effects not present in the atomic inversion, were discovered in the field entropy for the initial field state given by an statistical mixture of coherent states, namely the appearance of oscillations that indicate the collision of spots in phase space (Wigner function). Collisions (and its effects)  produced by different coherent states that initially were in a superposition, had been known already \cite{Banacloche}. However, effects by collisions of coherent states from statistical mixtures, to the best of our knowledge, are new.  


\begin{thebibliography}{Bi}
\bibitem{expl1} First the two atoms interact getting entangled and then one of them is introduced in the cavity to interact with the field.
\bibitem{expl2} First an atom passes through the cavity and interacts with the field via a dispersive Hamiltonian,  and, after the atom has exit the cavity another atoms interacts with the field.
\end{thebibliography}
\end{document}